\documentclass[runningheads,a4paper]{cmmr2023}
\usepackage{times}
\usepackage{amssymb}
\usepackage{graphicx}
\usepackage{subcaption}

\usepackage[hyphens]{url}
\usepackage{xcolor}
\newcommand{\keywords}[1]{\par\addvspace\baselineskip
\noindent\keywordname\enspace\ignorespaces#1}

\begin{document}

\mainmatter  

\title{Effects of Convolutional Autoencoder Bottleneck Width on StarGAN-based Singing Technique Conversion}
\titlerunning{Effects of Autoencoder Bottleneck Width on StarGAN-based STC}


%
%
\author{Tung-Cheng Su \and Yung-Chuan Chang \and Yi-Wen Liu \thanks{We thank the National Science and Technology Council of Taiwan for supporting this research under Grant No.~109-2221-E-007-094-MY3}}
%
\authorrunning{Tung-Cheng Su \and Yung-Chuan Chang \and Yi-Wen Liu}

\institute{Department of Electrical Engineering, National Tsing Hua University\\ \email{ywliu@ee.nthu.edu.tw}}

%

\maketitle

\begin{abstract}
Singing technique conversion (STC) refers to the task of converting from one voice technique to another while leaving the original singer identity, melody, and linguistic components intact. Previous STC studies, as well as singing voice conversion research in general, have utilized convolutional autoencoders (CAEs) for conversion, but how the bottleneck width of the CAE affects the synthesis quality has not been thoroughly evaluated. To this end, we constructed a GAN-based multi-domain STC system which took advantage of the WORLD vocoder representation and the CAE architecture. We varied the bottleneck width of the CAE, and evaluated the conversion results subjectively. The model was trained on a Mandarin dataset which features four singers and four singing techniques: the chest voice, the falsetto, the raspy voice, and the whistle voice. The results show that a wider bottleneck corresponds to better articulation clarity but does not necessarily lead to higher likeness to the target technique. 
Among the four techniques, we also found that the whistle voice is the easiest target for conversion, while the other three techniques as a source produce more convincing conversion results than the whistle.
\end{abstract}

\section{Introduction}

Singing voice conversion (SVC) is a task of converting 
prosodic features while retaining the linguistic content. The prosodic features to be converted can include singer identity, emotions, and  singing techniques.  Unlike speech conversion, the pitch  contour of the singing voice is usually unchanged in SVC so that the melody of the original voice is preserved. 

In recent years, many deep learning based methods of voice conversion (VC) have been shown to achieve state-of-the-art performance \cite{VCreview}, and several methods have also been applied to SVC \cite{StarGAN-VC,2019SVC,Pitchnet,2020STC,2021STC,fastSVC,DiffSVC}. Compared to speech, singing is rich in terms of the voicing techniques that singers can apply to enhance their expressiveness, such as to switch between their chest voice, falsetto, whistle voice, and so on. Thus, one's singing technique is a integral part of their singing performance \cite{VocalTimbre}, yet computer-based singing technique conversion (STC) is a less researched field compared to other SVC tasks. Previous works have applied the autoencoder (AE) for STC \cite{2020STC,2021STC} as well as other VC tasks \cite{StarGAN-VC,2019SVC,Pitchnet}; however, how the architecture of the AE affects the synthesized voice quality has not been thoroughly studied.

Therefore, in this study we focus on the \textit{bottleneck} of convolution autoencoders (CAEs) because it corresponds to the latent space representation of the features. Although the conversion process is operating within the latent space 
whose dimension is equal to the width of the bottleneck, the bottleneck architectures in existing SVC and STC models seem to be arbitrarily designed. To the best of our knowledge, few studies \cite{Bottleneck} focused on the effects of bottleneck architecture, and no study on bottlenecks was dedicated to STC or SVC 
in general, despite of AE's prevalence in the field. 
To gain about to STC and explore bottleneck architectures, we presently built a STC system based on StarGAN \cite{StarGAN}, and
experiments were conducted to compare the voice synthesis quality of STC with different bottleneck sizes.

The rest of this paper is organized as follows. Section 2 
describes our voice conversion system and introduces StarGAN. Then, details of experiments and evaluation methods are described in Sec.~3. 
Results are reported and discussed in Sec.~4, and 
conclusions are given in Sec.~5.

\section{Voice Conversion System Overview}
\begin{figure}
\centerline{\includegraphics[width=12.5cm]{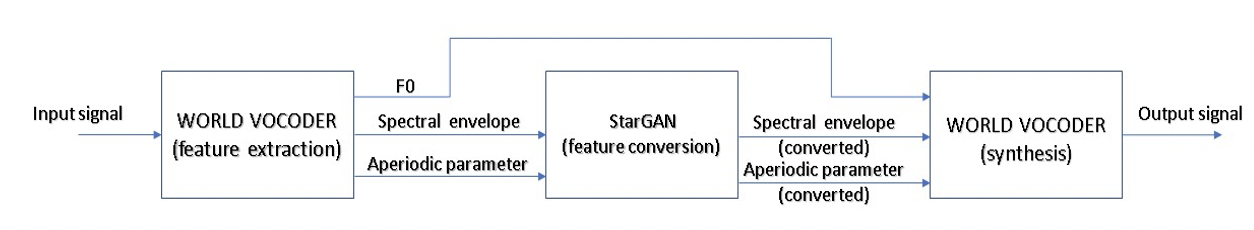}}
\caption{The overall system block diagram for the present research}
\label{system_block_diagram}
\end{figure}
Figure 1 shows the overall system diagram of this research. We adopted the WORLD vocoder \cite{WORLD} to represent the signal by three sets of features, namely the fundamental frequency (F0), the aperiodic parameters (AP), and the spectral envelope (SP).
The features were extracted every 5 ms.
We concatenated SP and AP into 56-dimension Mel Cepstral Coefficients (MCC) for the SP plus 4 dimensions for the AP. Putting this 60 dimension feature into the StarGAN\cite{StarGAN} model (illustrated in Fig.~2), we could then convert the signal in the vocoder domain. Figure 1 also shows that F0 was directly passed to the synthesizer in order to preserve the pitch of a singing voice. 

The usage of the WORLD vocoder might limit the synthesized audio quality compared to what could be achieved by neurovocoders, such as HiFi-GAN\cite{HifiGAN}. Nevertheless, we chose to work with WORLD for it enables us to separately consider F0 and other acoustic features, which suits our purpose of transforming the singing technique while maintaining the original F0.

\subsubsection{StarGAN}
StarGAN\cite{StarGAN} is a GAN-based model consisting of a generator and a discriminator. The generator adopts a convolutional autoencoder architecture as shown in Fig.~2, which could be divided into two stages. The first stage can be viewed as an encoding stage that downsamples features to the latent space; the second stage is a decoding stage that upsamples the latent feature back to the original space. In this research, the original space is the 60-dimensional 
WORLD vocoder output mentioned above. In Fig.~2, the middle section between the last layer of the encoder and the first layer of the decoder is referred to as the \textit{bottleneck}. 

The first layer before the downsampling layers uses a kernel of size 3x9 with a stride of 1. The output of each of the four downsampling layers are 30x200, 15x100, 5x100, and 1x100 respectively; their kernel sizes are 4x8, 4x8, 4x7, and 5x7 with corresponding stride of (2, 2), (2, 2), (3, 1), and (1, 1) respectively. For experiments, the model with only the first two, three, or four down/upsampling layers are used. The bottleneck size is thus controlled by the last downsampling layer, which is the same as the encoder's output. Upsampling layers mirror the downsampling layers with the same number of layers and their kernel size and stride. The last layer after the upsampling layers uses a kernel of size 7x7 with a stride of 1. 


In Fig.~2, the attribute vector encodes the singing technique of an audio file. Here, we define \textit{domain} as a set of audio files with the same attribute. Traditionally, a SVC model is only capable of performing conversion from one domain to another. In contrast, StarGAN achieves conversion between multiple domains with one single network. A key component of our proposed network is to
represent the attribute by several channels with the same height and width as the bottleneck, in a similar fashion as one-hot vectors\footnote{We specify one out of four possible target domains by setting one of four channels to be all 1.}. 
The target singing technique was thus informed to the decoder of CAE via the attribute channels.



The StarGAN is trained by minimizing the sum of three losses. The first is the adversarial loss, which makes the discriminator and generator work in an antagonistic fashion so that the generated features become more and more realistic. 
The second is the classification loss. The discriminator learns to classify WORLD vocoder features in the training set, while the generator aims to convert the features so that the classifier would put them into the target category. The third is the reconstruction loss. It forces the generator 
to reconstruct original features when given the original attribute vector. A lower reconstruction loss indicates less information loss in the bottleneck. In STC, lower reconstruction loss often indicates higher articulation clarity. Empirically, we observed that the width of the bottleneck had a significant impact on the reconstruction loss. This observation motivated us to conduct the experiments described next.
\begin{figure}
\centerline{\includegraphics[width=12.6cm]{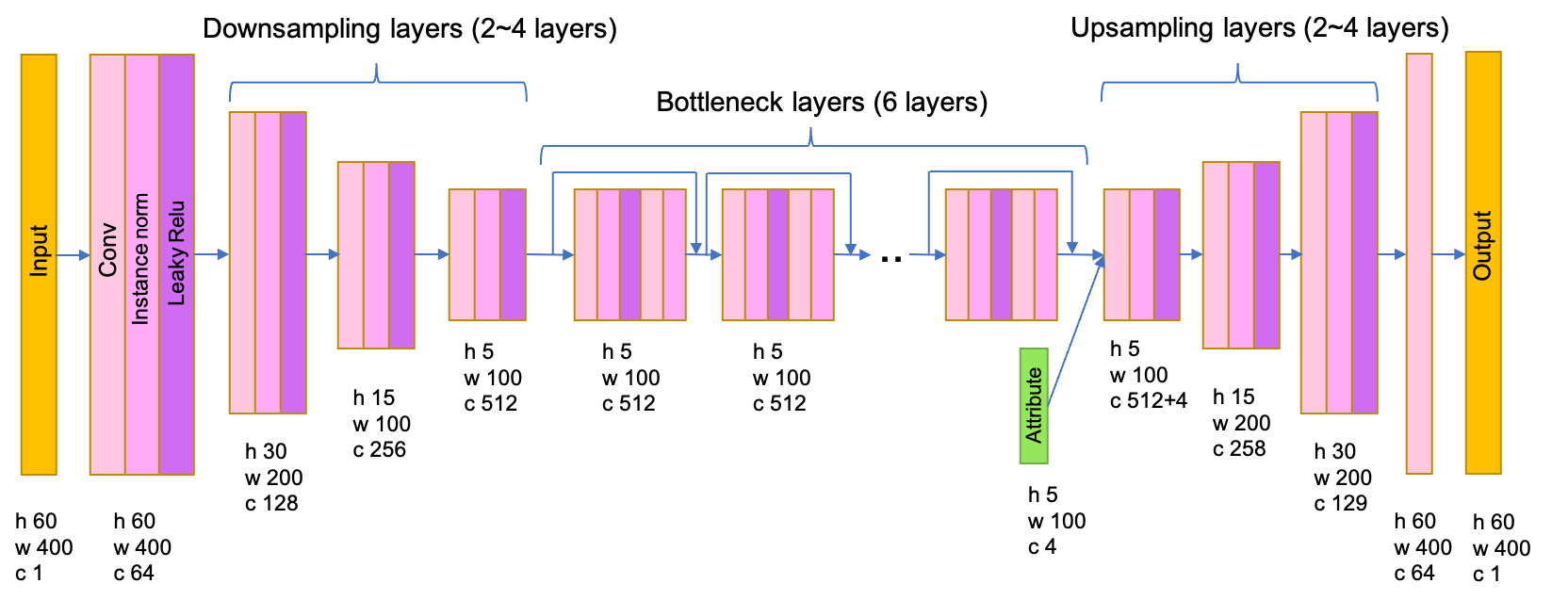}}
\caption{The proposed convolutional autoencoder architecture for StarGAN. 
The figure shows a version of the model with three down/upsampling layers. The three downsampling layers have kernel sizes 4x8, 4x8, and 4x7, with strides of (2, 2), (2, 2), and (3, 1) respectively; the upsampling layers' settings mirror those of the downsampling layers.
}
\label{model}
\end{figure}

\section{Experiments}
In this section, we describe the dataset, the network training strategies, the bottleneck architecture that was varied, and the evaluation metrics for this study.
\subsection{Dataset}
While existing datasets, such as VocalSet \cite{VocalSet}, already featured diverse singing techniques and were used in previous STC studies \cite{2020STC,2021STC}, we decided to collect a new dataset from scratch so as to focus on the singing techniques that are common in Chinese Mandarin pop singing.
Our dataset contains non-parallel singing voice of four singing techniques, namely the chest voice, falsetto voice, whistle voice, and raspy voice. Two male and two female singers were recruited. Each singer sang 
Chinese Mandarin pop songs in their preferred techniques while they were instructed to maintain the same pitch range across different techniques, except for the whistle voice. Since the techniques were constrained by the singers' preferences, not all techniques were successfully recorded from all singers. In the end, the chest voice was sung by all four singers for a total of 53 minutes, the falsetto voice was sung also by all four singers for a total of 51 minutes, the whistle voice was sung by one female singer for 20 minutes, and the raspy voice was sung by one male singer for 7 minutes.


 The audio was recorded with a large diaphragm condenser microphone\footnote{Sontronics STC-2, without the built-in high-pass filtering or -10dB passive attenuation}
 and sampled at 48 kHz in a vocal booth to approximate the recording environments of pop music vocals. The recordings were cut phrase by phrase afterwards, with each phrase being between 5 to 12 seconds. The audio data were then re-sampled to 16 kHz for STC experiments.

\subsection{Training configurations}
 For data augmentation, the model randomly selected 400 continuous frames (2 seconds) from the training set each time. We used the Adam optimizer \cite{Adam} for training with \(\beta_1=0.5\) and \(\beta_2=0.999\) for all the models. Each model was trained for 250,000 iterations with \(10^{-4}\) learning rate at the start and decays for the last 100,000 iterations. To optimize training quality, we updated the discriminator once for every three generator updates.

\subsection{Bottleneck configurations}
Our experimental design aims to investigate the influence of bottleneck width on singing voice technique conversion performance. Hence, we formulated three different bottleneck sizes, $15\times 256$, $5\times512$, and $1\times 1024$ (features x channels). To only changes bottleneck sizes and not other CNN settings, downsampling/upsampling layers are added or eliminated for different sizes; as illustrated in Fig.~2, these three configurations correspond to encoders with two, three, and four downsampling layers, respectively.


\subsection{Subjective Evaluation}
The subjective evaluation test consisted of three listening tasks, and 27 participants were recruited. 

\subsubsection{Bottleneck Comparison}
We compared the conversion performance and articulation clarity of the synthesized voice produced by three different bottleneck widths across four distinct source techniques. Eight listening comparison tests were created. In each test, participants were provided with source and target audio samples beforehand for familiarization purposes. Then, they were asked to {rank} the audio conversion results of three different bottlenecks in terms of conversion performance and articulation clarity. The ranking was then given a score from 1-3, with the best receiving 3 points, 2 points for second-best, and 1 point for the worst. 

\subsubsection{Likeness to the target}
The performance of the multi-domain STC model was evaluated in terms of likeness to the target technique after a subject listened to the transformed audio. The model with 3 encoding layers was chosen for evaluation. Similar to the bottleneck comparison experiment, we provided the participants with source and target audio files for familiarization, but asked them to rate the transformed audio on a scale of 1-5, 
where 5 means most similar to the timbre of the target file, and 0 means most similar to that of the source file.
For this part of the experiment, we performed pitch shifting when the whistle voice was involved in the conversion so the
target pitch range sounded natural to the intended singing technique. 

\subsubsection{Sound Quality}
The final part of subjective evaluation aims to assess degradation in the sound quality after STC. To achieve this, we selected four audio files (C, F, W, R) and processed them by analysis-then-synthesis using the WORLD vocoder; the same audio files were also subjected to STC (C2F, F2W, W2R, R2C) so their sound quality could be evaluated.
\section{Results and Discussion}

\begin{figure}
\begin{subfigure}{.5\textwidth}
    \centering
    \includegraphics[width=0.9\linewidth, height = 3cm]{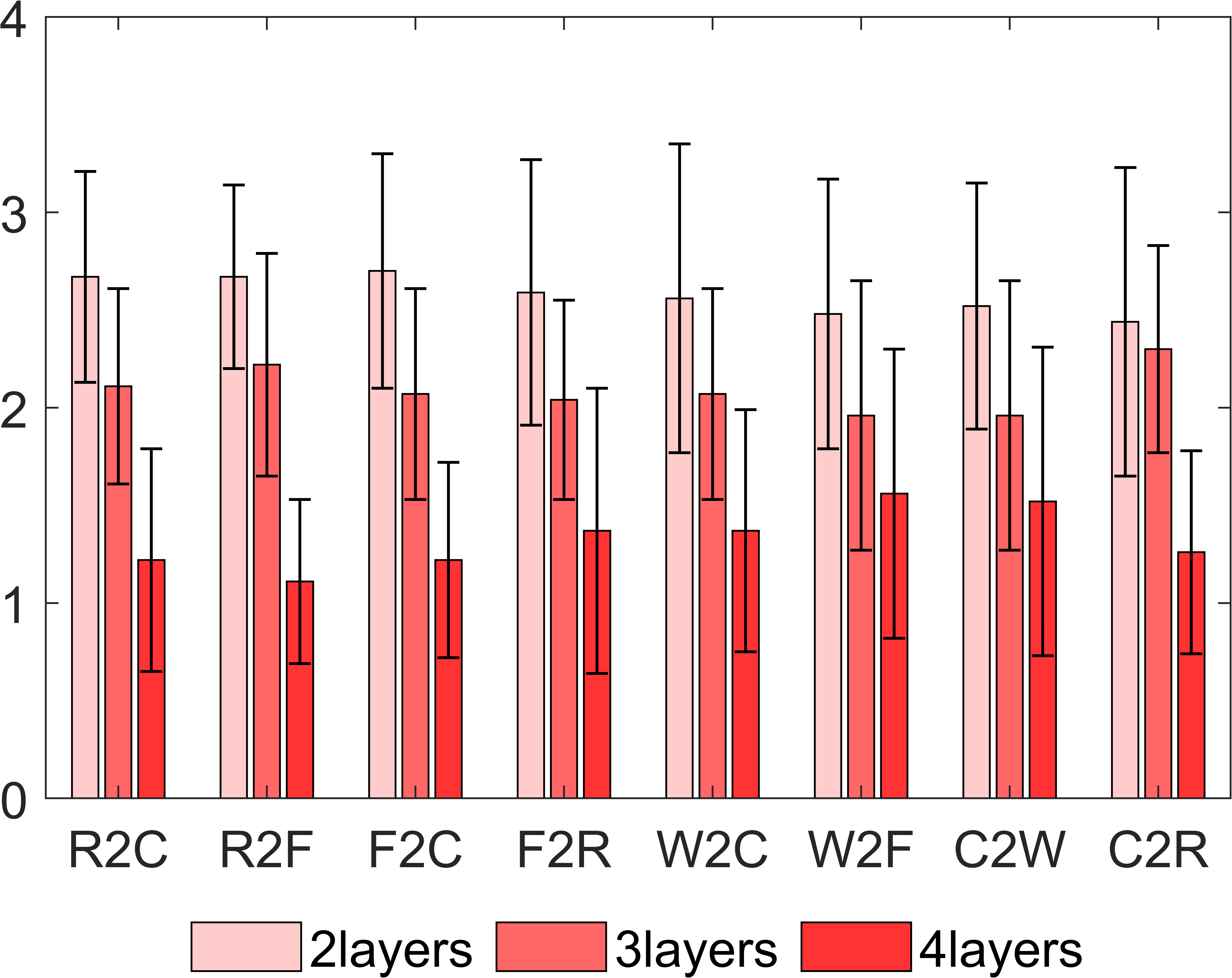}
    \caption{Articulation clarity comparison test}
    \label{clarity_bottlenecks}
\end{subfigure}
\begin{subfigure}{.5\textwidth}
    \centering
    \includegraphics[width=0.9\linewidth, height = 3cm]{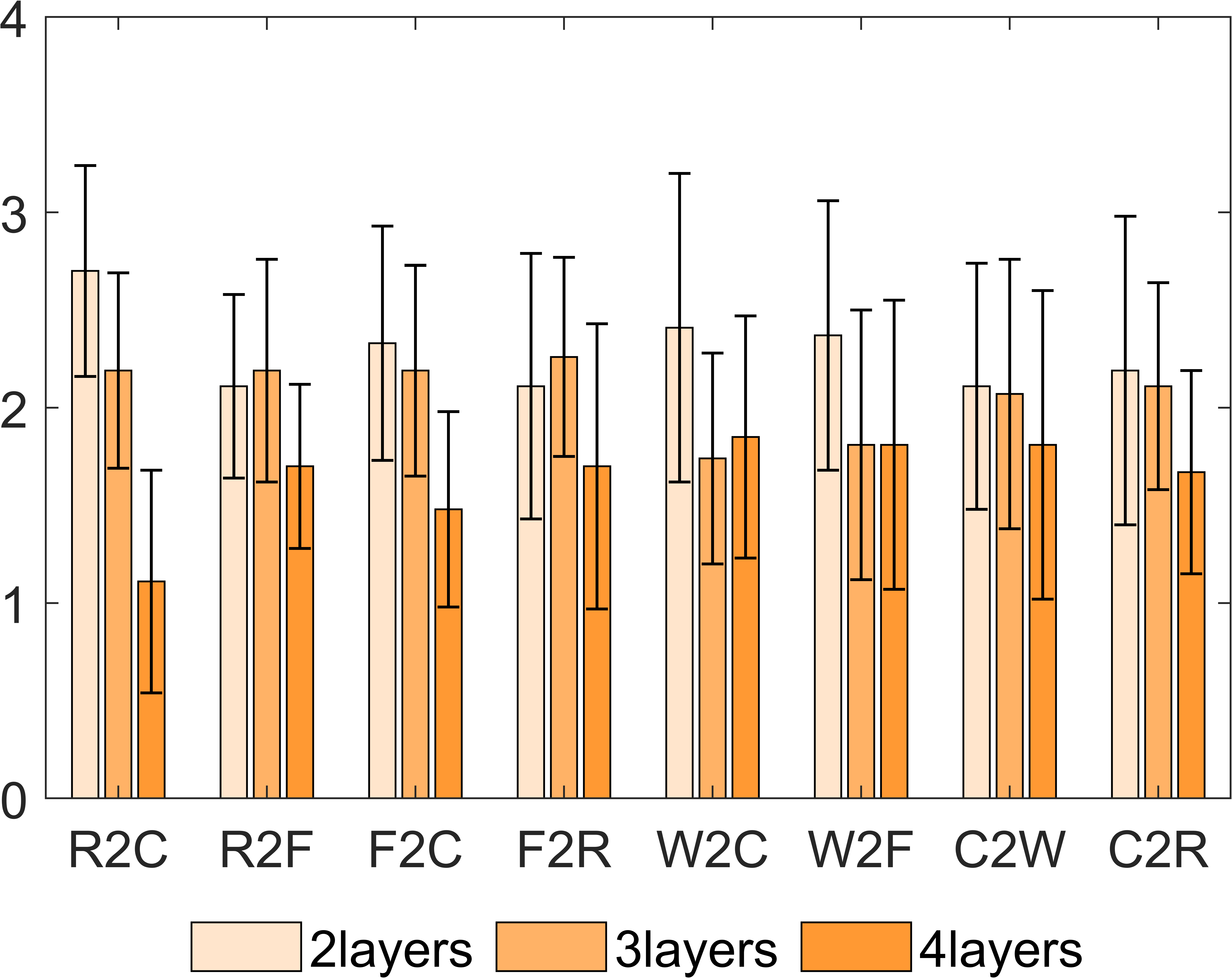}
    \caption{Conversion performance comparison test}
    \label{conversion_bottlenecks}
\end{subfigure}
\caption{For both comparison tests, each audio was scored on a scale of 1-3, with the best receiving 3 points, 2 points for second-best, and 1 point for the worst. 
C = chest voice, F = falsetto, W = whistle voice, and R = raspy voice. The error bar represents one standard deviation.}
\label{bottleneck_comparisons}
\end{figure}

\begin{figure}
\begin{subfigure}{0.5\textwidth}
    \includegraphics[width=0.8\linewidth, height = 2.7cm]{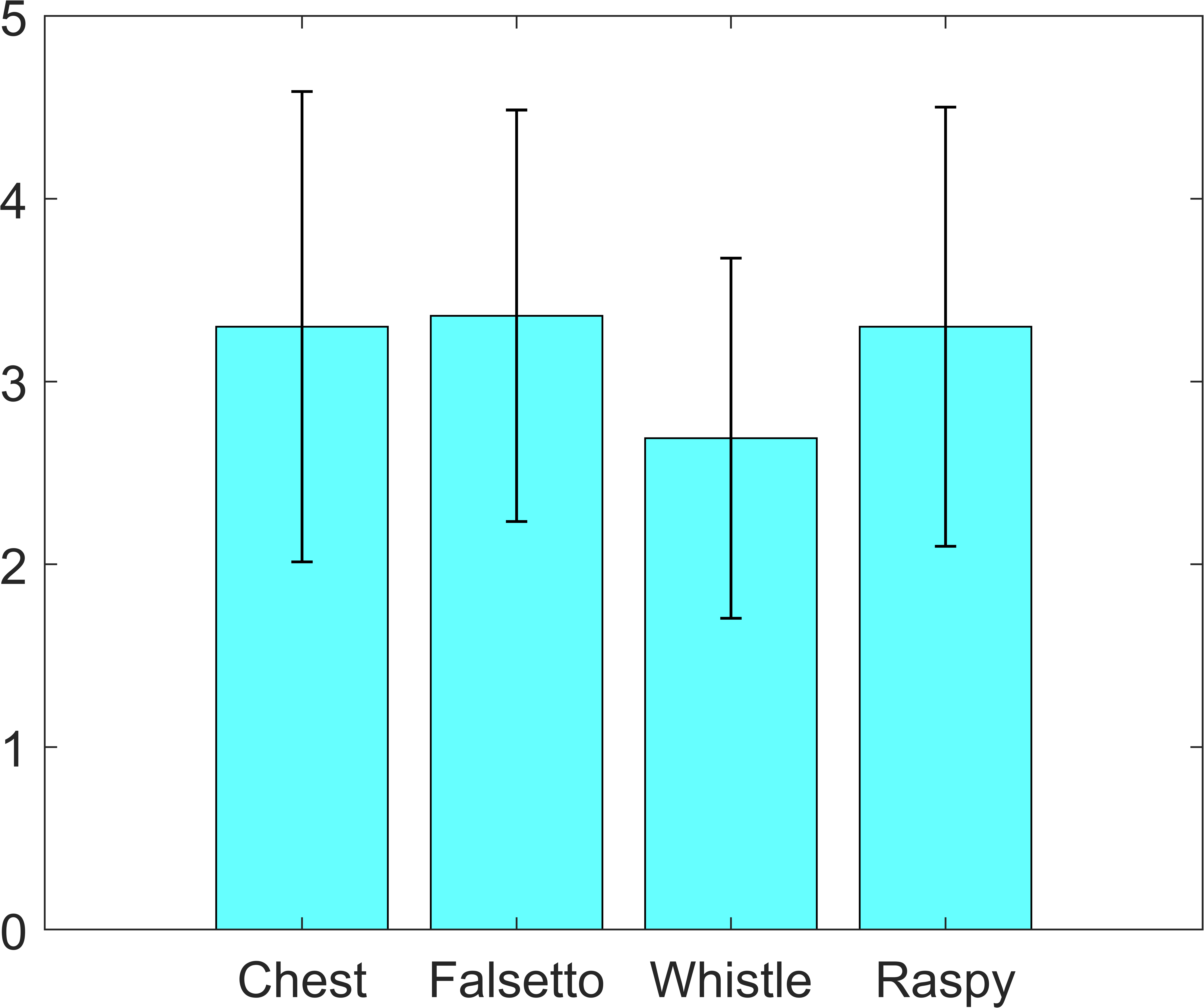}
    \caption{Likeness score of each source technique}
    \label{source_MOS}
\end{subfigure}
\begin{subfigure}{0.5\textwidth}
    \includegraphics[width=0.8\linewidth, height = 2.7cm]{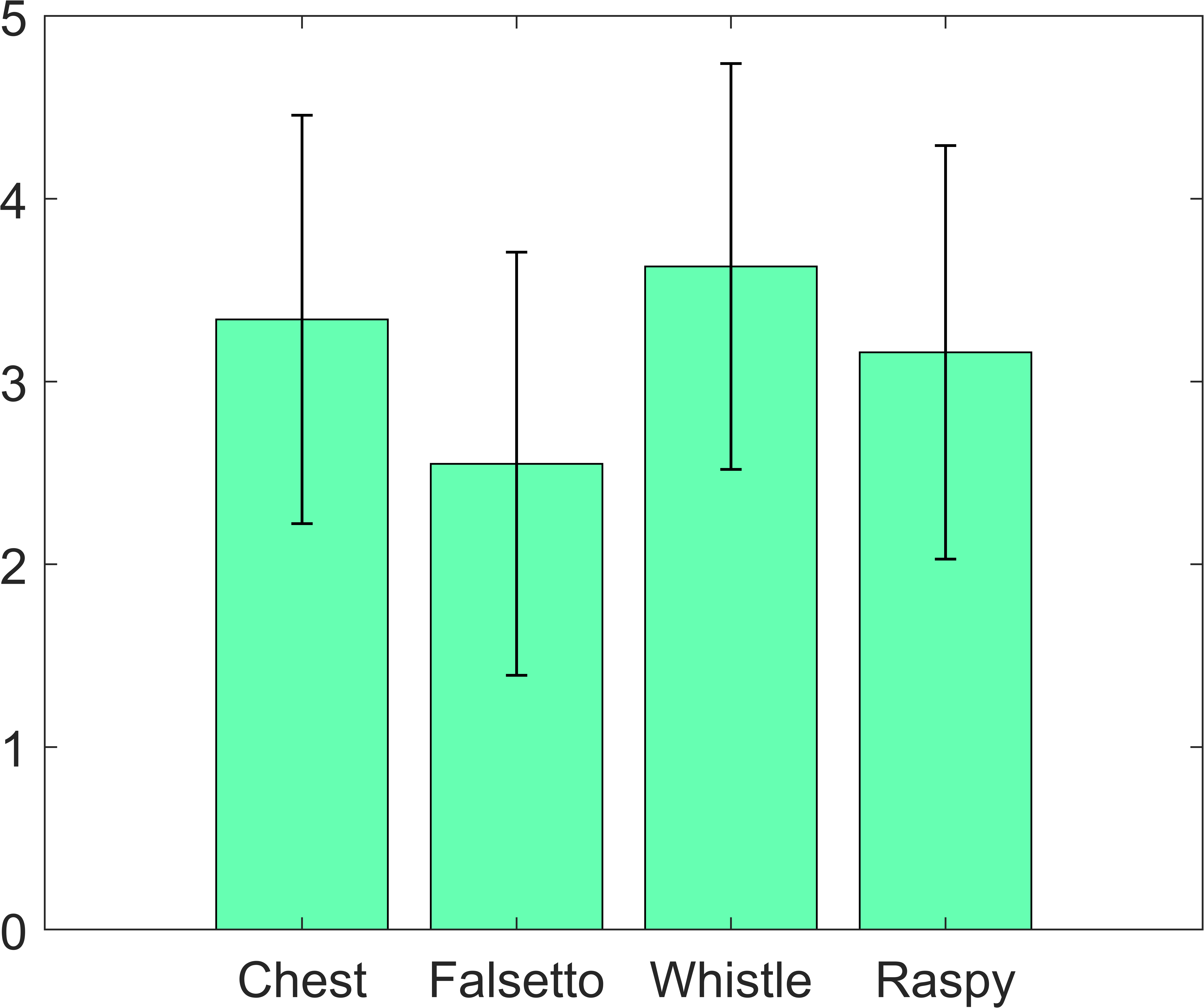}
    \caption{Likeness score of each target technique}
    \label{target_MOS}
\end{subfigure}
\begin{subfigure}{\textwidth}
    \centering
    \includegraphics[width=0.5\linewidth, height = 3cm]{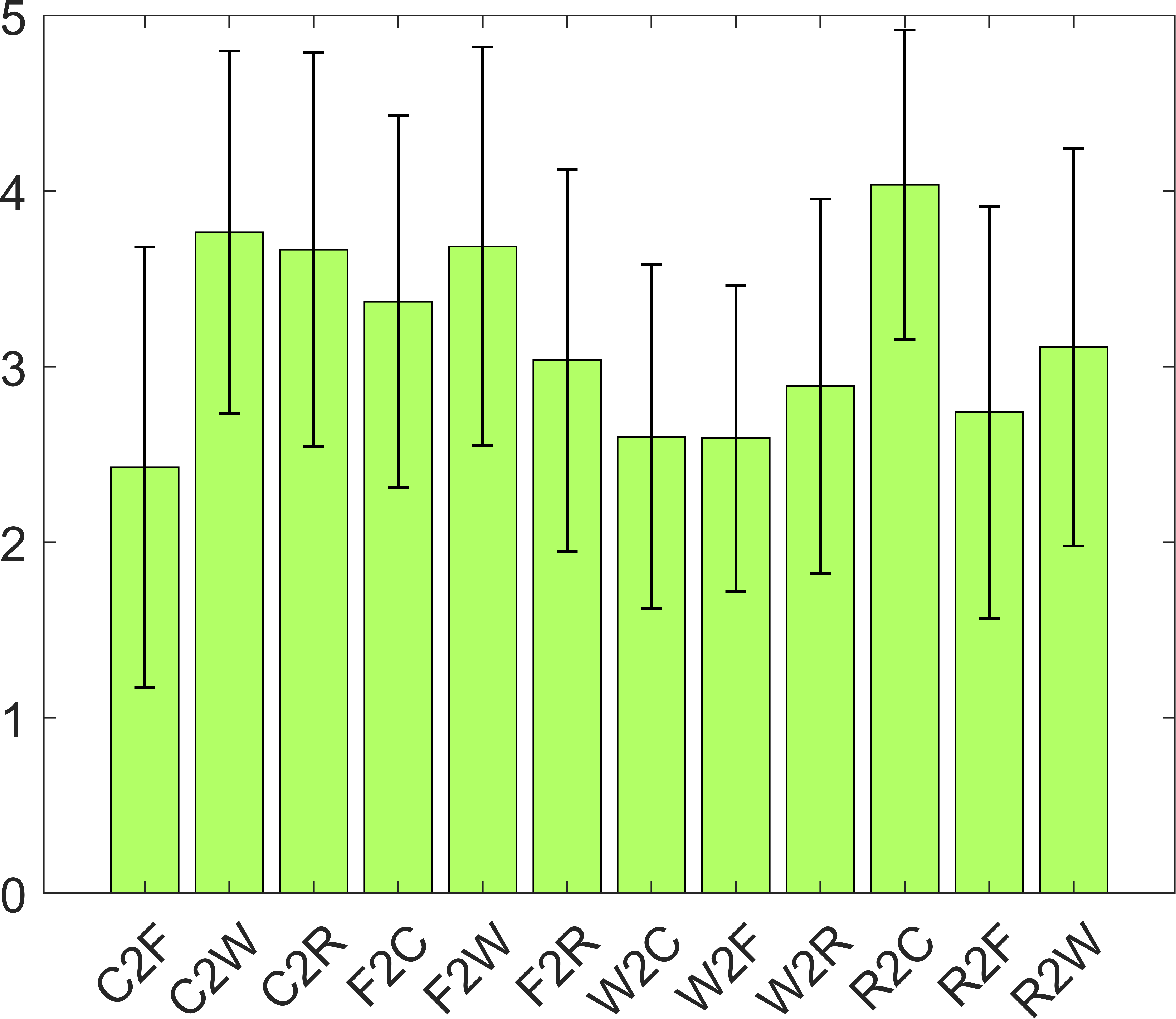}
    \caption{Likeness score of each source-target combination}
    \label{path_MOS}
\end{subfigure}
\caption{For the likeness tests, 5 means most similar to the timbre of the target, and 0 means most similar to that of the source. 
C = chest voice, F = falsetto, W = whistle voice, and R = raspy voice. The error bar represents one standard deviation.}
\label{conversion_MOS}
\end{figure}


Figure \ref{bottleneck_comparisons} shows the mean and standard deviation of the 3-point scores for 8 different source-target combinations under three different widths of the CAE bottleneck. 
The results indicate that, in general, the widest bottleneck (which is 15 and corresponds to two encoding layers) produced the most clearly articulated speech. This makes sense, since a wide bottleneck encodes the input information into a higher-dimensional latent space and thus preserves more complete information about the voice content. A narrower bottleneck, in contrast, encodes information into a lower-dimensional latent space, making it more difficult to reconstruct the audio.

However, better articulation clarity was not always accompanied with a better conversion performance. Particularly, for conversion between raspy and falsetto voices (i.e., R2F and F2R), the results obtained with three downsampling layers in the encoder were slightly superior to those obtained with two downsampling layers. Also, while the system with four downsampling layers produced poor articulation clarity for the W2C transformation, it slightly outperformed the system with three encoding layers in terms of conversion performance. These findings suggest that the selection of an optimal bottleneck size is critical for singing voice techniques conversion, and best setting might depend on the intended source-target combination.

Figure \ref{conversion_MOS} summarizes our evaluation of the system in terms of the timbral likeness to the target technique after conversion.
In (a), the average results of four \textit{source} techniques are shown, and (b), four \textit{target} techniques; the average results of all source to target pair are shown in (c).
The results indicate that the conversion of whistle voice to other techniques had limited success (with mean score $< 3.0$). However, the transformation from other techniques to whistle voice was effective, with a mean score of {3.63}. 
This can be attributed to the unique timbre and the high pitch range of whistle voice, which were probably difficult to remove and easy for the listeners to recognize. The conversion of falsetto voice as a source has yielded satisfactory results, but achieved significantly poorer score as a target. Chest and raspy voices produced comparable results, regardless of them being utilized as source or target technique.

 The mean opinion score (MOS) on a five-point scale was $3.55\pm1.04$ for WORLD vocoder round-trip,
 and $2.80\pm1.30$ after STC. Although the mean score decreased reasonably by $0.75$ due to STC, several limitations of WORLD vocoder were noted in this research. First,
 we observed that WORLD encoding-decoding produced some cracking or breaking sound when we tested on the raspy voice. 
 We suspect that the WORLD vocoder might have been optimized to handle monophonic sounds, whereas a raspy voice can have multiple concurrent fundamental frequencies, causing the vocoder to misinterpret the data. 
 Additionally, we observed that the aspiration that was salient in the whistle voice could cause errors in voiced/unvoiced classification and thus pose a challenge for the vocoder-domain processing. To summarize, future fine-tuning of the vocoder should be warranted for improving the quality of STC.

\section{Conclusion}
In this research, we created a singing voice technique dataset that includes chest voice, falsetto, raspy voice, and whistle voice.
The dataset was adopted to train a multi-domain singing technique conversion model. 
We found that the size of CAE's bottlenecks affected the clarity of pronunciation and the likeness to the target technique after conversion, and the optimal size might depend on the intended source-target combination. Furthermore, we noted several audible defects when handling raspy or whistle voices with the WORLD vocoder, which ultimately limited the audio quality of STC. In the future, we hope to continue improving the audio quality of STC and create different ways of vocal music production for amateurs and professionals to use.

\end{document}